\documentclass[12pt]{article}

\newfont{\msbm}{msbm10}
\usepackage{yfonts}
\usepackage[T1]{fontenc}
\usepackage{a4wide}

\begin{document}

\title{Unitary evolution and uniqueness \\ of the Fock quantization in flat cosmologies\\ with compact spatial sections}

\author{L. Castell\'o Gomar\thanks{Email:laura.castello@iem.cfmac.csic.es}\\
Instituto de Estructura de la Materia, IEM-CSIC, \\
Serrano 121, 28006 Madrid, Spain 
\and 
J. Cortez\thanks{Email:jacq@ciencias.unam.mx}\\
Departamento de F\'{\i}sica, Facultad de Ciencias, \\
Universidad Nacional Aut\'onoma de M\'exico,\\ M\'exico D.F. 04510, Mexico \\
Departamento de F\'{\i}sica,\\ Universidad Aut\'onoma Metropolitana-Iztapalapa, \\
San Rafael Atlixco 186, M\'exico D.F. 09340, Mexico
\and
D. Mart\'{\i}n-de~Blas\thanks{Email:d.martindeblas@uandresbello.edu}\\
Departamento de Ciencias F\'{\i}sicas,\\ Facultad de Ciencias Exactas, 
Universidad Andr\'es Bello, \\ Av. Rep\'ublica 220, Santiago 8370134, Chile
\and
G. A. Mena~Marug\'an\thanks{Email:mena@iem.cfmac.csic.es}\\
Instituto de Estructura de la Materia, IEM-CSIC, \\
Serrano 121, 28006 Madrid, Spain
\and 
J. M. Velhinho\thanks{Email:jvelhi@ubi.pt}\\
Departamento de F\'{\i}sica,  Universidade da Beira Interior, \\ 
R. Marqu\^es D'\'Avila e Bolama, 6201-001 Covilh\~a, Portugal}

\maketitle

\begin{abstract}
We study the Fock quantization of scalar fields with a time dependent mass in cosmological scenarios with flat compact spatial sections. This framework describes physically interesting situations like, e.g., cosmological perturbations in flat Friedmann-Robertson-Walker spacetimes, generally including a suitable scaling of them by a background function. We prove that the requirements of vacuum invariance under the spatial isometries and of a unitary quantum dynamics select (a) a unique canonical pair of field variables among all those related by time dependent canonical transformations which scale the field configurations, and (b) a unique Fock representation for the canonical commutation relations of this pair of variables. Though the proof is generalizable to other compact spatial topologies in three or less dimensions, we focus on the case of the three-torus owing to its relevance in cosmology, paying a especial attention to the role played by the spatial isometries in the determination of the representation.
\end{abstract}


\section{Introduction}
\label{intro}

The construction of a quantum theory to describe a classical system is a process plagued with many ambiguities. Generically, the correspondence is not one-to-one. What is more important, the quantum physics depends on the choices made in the steps affected by those ambiguities. The specification of a unique quantization must then either rest on the confrontation of the predictions with experiments, or be achieved by appealing to other kinds of criteria, usually related to symmetries of the system or to the behavior of the quantum states. The problem is especially relevant in cosmology, both because the windows for quantum effects in cosmological observations are certainly narrow (if any), and because one cannot really select the best candidate for a quantum model of the universe by performing an indefinite number of repeated measurements (in copies of the same state), since we can only observe the universe in which we live.

Even if one identifies a classical system with a specific algebra of observables, attained by first selecting a set of appropriate variables for the system and constructing out of them the algebra of functions which we declare of interest, its quantization is still not unique. In general, there will exist representations of that algebra which are not equivalent (that is, which are not related by a unitary transformation). This problem appears not just for generic systems with complicated phase spaces or interactions, but also for the simplest systems with a linear behavior. In the case of standard quantum mechanics, when the number of degrees of freedom is finite, the ambiguities in the representation can be removed by imposing additional requirements. For the Weyl algebra corresponding to the free particle, one usually demands that the representation be irreducible, unitary and strongly continuous. Then, the Stone-von Neumann theorem guarantees that the representation is unique (up to unitary transformations) \cite{Vonneu}. A similar result does not exist when one considers systems with fields, which have an infinite number of degrees of freedom. The ambiguities persist even if one takes advantage of the linear structures and restricts the consideration to representations of the Fock type, where a concept of particle and vacuum are available (either with a genuine physical interpretation or just as mathematical entities). The possible choices of Fock representation which are physically different are still infinite, and correspond to non-equivalent choices of a vacuum \cite{wald}.

For fields propagating in highly symmetric spacetimes, the symmetries of the background can be employed as criteria to determine the quantization \cite{wald,a-m,jackiw}, asking e.g. that all the quantum structures incorporate those symmetries. For instance, this happens in Minkowski spacetime, where the vacuum can be selected by demanding Poincar\'e invariance \cite{wald}. But, for more general spacetimes, no generic criterion exists that specifies the Fock representation. In generic situations, there is not enough spacetime symmetry to fix it. Therefore, supplementary or alternate requirements are needed in order to pick out a unique equivalence class of Fock representations. Frequently, the situation found in cosmology is that the spatial sections still present a high degree of symmetry (at least in a certain approximation), but the stationarity is lost owing to the universe expansion (or contraction). In this framework, it seems natural to adhere to the remaining spatial symmetries, demanding that they are naturally included in the quantum theory, and replace the criterion of time symmetry with the closest possible one, namely, with the requirement of a unitary evolution. Unitarity will guarantee a standard probabilistic interpretation in the quantum theory, though the loss of time symmetry will make the vacuum change dynamically. Actually, the combined criterion of invariance under the spatial symmetries and a unitary evolution has been successfully employed recently in the selection of a unique quantization for cosmological models.

Apart from the choice of representation, there is an additional ambiguity that arises naturally in the construction of a Fock description for fields in cosmological scenarios. The non-stationarity of the spacetime leads to the obvious possibility of absorbing part of the time dependence of the field via its scaling by a function that depends only on the background. This is so irrespective of whether the spacetime in which the propagation takes place is a true physical background \cite{mukhanov,b-d}, an effective background (emerging from an effective description of the system, for instance incorporating some quantum modifications at an effective level \cite{AKL,PRD85,PRD86,PRD88}) or an auxiliary background (like, e.g., in dimensional reductions of gravitational systems in General Relativity, using the presence of Killing symmetries \cite{Gowdy,PRD72,unit-gt3}). Scalings of this type are found in many circumstances when considering fields or cosmological perturbations around Friedmann-Robertson-Walker (FRW) spacetimes, for instance, like in the case of test fields, or scalar and tensor perturbations, including their description in terms of gauge invariants, as with the Mukhanov-Sasaki variables \cite{muksasaki}. These scalings render the field equations in the form of those corresponding to a Klein-Gordon (KG) field in a static (auxiliary) spacetime. The time dependence still shows up via the appearance of a varying mass term. Besides, usually there is no friction term in the scaled equation {(or one can neglect it when considering short scales in the field behavior). In particular, it was demonstrated recently that, for KG equations in conformally ultrastatic spacetimes, one can always find an adequate scaling which, properly combined with a change of time, removes the term proportional to the first time derivative in the field equations \cite{generalization}.} The scaling of the field configuration can always be regarded as part of a time dependent canonical transformation in the system, which is natural to consider as a local and linear transformation in order not to spoil this type of properties in the field system. In such a transformation, the field momentum gets the inverse scaling of that of the field configuration [to maintain the canonical commutation relations (CCR's)] and, in addition, may admit an extra contribution linear in the field configuration, multiplied by a function of time. Any of these canonical transformations leads to a new canonical pair for the field, but also alters the dynamics, since the change is time dependent. The quantum description can be made by adopting any of these canonical pairs for the field, introducing a new kind of ambiguity with infinitely many possibilities.

As we have briefly commented, the criterion of invariance under the spatial symmetries of the classical spacetime, together with the unitary implementation of the dynamics in the quantum realm, has been employed in cosmological systems to remove the ambiguity in the Fock quantization. Indeed, this criterion has shown useful not only to select a Fock representation, but also to determine a unique canonical pair for the field among all those that can be reached by means of linear canonical transformations which depend on time. Note that, since these transformations modify the dynamics, the fact that the evolution is unitary is intimately related to the canonical pair adopted. The criterion was first applied to the case of the Gowdy models \cite{PRD72,unit-gt3,ccmv1,PRD76,PRD75,BVV2,CQG25}: reductions of General Relativity with two spatial Killing vectors and compact spatial sections. They describe gravitational waves propagating in cosmological spacetimes with compact universes. In the case of linearly polarized waves, the system admits a description in terms of a KG field (with time dependent mass) in an auxiliary, dimensionally reduced stationary spacetime. The spatial sections of this auxiliary background can be either isomorphic to the circle, $S^1$, or to the sphere, $S^2$, depending on the spatial topology of the Gowdy model (which can be that of a three-torus, of a three-sphere, or of a three-handle, $S^2\times S^1$). The conclusions about the validity of the proposed criterion for the choice of a unique Fock quantization in the Gowdy models have been lately extended to the case of backgrounds with spatial sections isomorphic to $d$-dimensional spheres, with $d\leq 3$ \cite{CMV8,PRD79,CMV83,JCAP10}, and even more recently to a general compact topology in three or less spatial dimensions \cite{CMOV28,gr-qc}. Although this generalization proves that the uniqueness holds in any case with compact spatial sections, we will focus here on the case when these sections have the topology of a three-torus. The relevance of this case is clear since it describes flat universes, which is precisely the favored scenario for the universe according to observations. Moreover, the discussion for fields in spacetimes with generic compact topology \cite{CMOV28,gr-qc} is obscured by the fact that there may be no clear geometric interpretation of the considered symmetry group, in the most general case. This interpretation is neat in the case of the three-torus. Besides, there exist certain peculiarities which motivate the mathematical interest in the analysis of the three-torus. Namely, since the group of isometries of the three-torus is an Abelian compact group (as in the $S^{1}$ case \cite{PRD79}), its irreducible (unitary) representations are one-dimensional and defined over complex vector spaces \cite{brocker}. This introduces certain subtleties in the characterization of the Fock representations which are compatible with those symmetries. These subtleties arise because the complex representations of the symmetry group must be combined suitably so that, at the end of the day, one deals exclusively with real scalar fields. In order to fully take into account this issue, and show explicitly how those complex representations combine, a departure from the general approach followed for other topologies in the literature is adopted \cite{CMV8,CMV83}. Explaining the characterization of the symmetric representations will be one of the main goals of this work.

The paper is organized as follows. In Sec. \ref{s2} we start by considering a KG field with time dependent mass in a static spacetime with flat spatial sections of three-torus topology. We briefly present the classical system and the standard procedure to introduce a Fock quantization of its associated phase space. Then, in Sec. \ref{s3} we prove that our combined criterion of a) invariance under the isometries of the three-torus and b) unitary implementation of the dynamics, selects a unique Fock representation, up to unitary transformations. This unitary class of representations includes the one which would be naturally adopted if the field had vanishing mass. Our proof contains a detailed discussion of the characterization of the representations which are invariant under the symmetries of the three-torus, with a careful treatment of the symmetry transformations and the consistency conditions coming from the reality of the field. We go beyond this result of uniqueness in Sec. \ref{s4}, where we analyze time dependent canonical transformations arising from a scaling of the field configuration, and demonstrate that only one of all those transformations is compatible with our requirements of spatial symmetry invariance and unitary evolution: the trivial one. Finally, we present our conclusions in Sec. \ref{s5}. The discussion in this work follows {lines} of arguments similar to {those presented in Refs. \cite{gr-qc,3torus}}, to which we refer the reader for further details about the analysis.

\section{Klein-Gordon field with time dependent mass}
\label{s2}
\subsection{The classical model}
\label{s21}

Before proving that the criterion that we put forward indeed selects a unique class of unitarily equivalent Fock representations for a linear scalar field with time dependent mass in a flat spacetime with compact sections, let us start by describing the classical set-up of our theory.

We consider a real scalar field $\varphi$ defined on a flat spacetime whose spatial sections have the topology of a three-torus, $T^3$.
These sections are equipped with the standard spatial metric of the three-torus, $h_{ij} (i,j=1,2,3)$. The field is subject to a linear equation of KG type:
\begin{equation}
\ddot{\varphi}-\Delta\varphi + s(t)\varphi =0,
\end{equation}
where the dot stands for the time derivative, $\Delta$ is the Laplace-Beltrami (LB) operator
associated with the three-torus metric $h_{ij}$, and $s(t)$ can be interpreted as a time dependent mass. In principle, $s(t)$ can be any time function, and only later in the discussion we will impose on it some extremely mild conditions about its derivatives. On the other hand, the time domain in which the KG field is defined can be any arbitrary connected real interval $\hbox{\msbm I} \subset \hbox{\msbm R}$. We do not impose that $\hbox{\msbm I}$ be the real line, not even that it be unbounded. No specific form for the interval $\hbox{\msbm I}$ is assumed. This is important for the applications of our results to situations in which the field description is effective, since the validity of the effective spacetime geometry can be restricted just to a certain {time} interval. More generally, the domain  $\hbox{\msbm I}$ might even be just a union of connected components. In such cases, the restriction to one of the components is sufficient to achieve the same uniqueness results.

This equation can be obtained in very different cosmological models. An important class of systems for which it has a major relevance are scalar fields propagating in non-stationary, cosmological spacetimes, like e.g. the case of matter fields in inflationary backgrounds \cite{mukhanov,PRL49,LR314}. Another class are cosmological perturbations around FRW spacetimes. According to our comments in Sec. \ref{intro}, a suitable {choice of the time parametrization and a} time-dependent scaling leads the field equations of those systems to a KG equation of the above type \cite{CMV83,JCAP10}. In this cosmological context, the case of flat spatial sections that we study here is the most interesting one owing to its potential applications to describe situations in the observed universe.

Clearly, this dynamical equation is invariant under the group of isometries of the three-torus, since the LB operator is defined in terms of the standard metric on $T^3$. To incorporate and analyze the role of these isometries, we consider the composition of rigid rotations in each of the periodic spatial directions $\theta_i$ that diagonalize the metric $h_{ij}$,
\begin{equation}
T_{\alpha_i} : \theta_i \to \theta_i+\alpha_i,\quad \forall\alpha_i\in S^{1}.
\end{equation}
Here, $\alpha_i$ is the angle parameter that provides the rotation in the direction $i$ ($i=1,2,3$). We will call $T_{\vec{\alpha}}$, with $\vec{\alpha}=(\alpha_1,\alpha_2,\alpha_3)$, the transformation obtained by composing the corresponding rotations.

The canonical phase space $\Gamma$ of the field system is obtained from the Cauchy data at a reference time $t_0\in \hbox{\msbm I}$, namely $\{(\varphi, P_\varphi)\} = \{(\varphi_{|t_0}, \dot{\varphi}_{|t_0})\}$, equipped with a symplectic structure $\Omega$ that amounts to the canonical Poisson brackets $\{\varphi(\vec{\theta}),P_\varphi(\vec{\theta'})\} = \delta^{3}(\vec{\theta}-\vec{\theta'})$, with $\delta^{3}(\vec{\theta})$ being the Dirac delta on $T^3$ and $\vec{\theta}$ the spatial point with coordinates $\theta_i$. Note that we assume that the Hamilton equation for the field momentum is $P_{\varphi}=\dot{\varphi}$.\footnote{There is no problem with the density weight of the field momentum in this equation, since the standard metric of the three-torus has a unit determinant.}

As a result of the periodicity on the spatial coordinates $\theta_i$, one can decompose the field $\varphi$ (and its momentum) in an expansion in Fourier modes. Note that these modes are eigenfunctions of the LB operator. The field decomposition using these complex eigenfunctions takes the form
\begin{equation}
\varphi(t,\vec{\theta})=\frac{1}{(2\pi)^{3/2}}\sum_{\vec{m}}\hbox{\frakfamily q}_{\vec{m}}(t)\exp\{i(\vec{m}\cdot\vec{\theta})\},
\end{equation}
where $\vec{m}$ is the tuple of integers $(m_1, m_2, m_3)$ (i.e., $m_i\in \hbox{\msbm Z}$, for $i=1,2,3$), and we have introduced the notation $\vec{m}\cdot\vec{\theta}=\sum_{i} m_i\theta_i$. Recalling that the field is real, we find that the coefficients of the expansion are subject to the reality conditions
\begin{equation}
\hbox{\frakfamily q}_{-\vec{m}}(t)=[\hbox{\frakfamily q}_{\vec{m}}(t)]^*,
\end{equation}
where the symbol $*$ denotes complex conjugation.

This complex decomposition is well adapted to the three-torus symmetries. The Fourier modes are eigenfunctions of all the transformations $T_{\vec{\alpha}}$.  Moreover, each tuple $\vec{m}$ provides a different, inequivalent irreducible representation of the isometry group. In particular, we straightforwardly see that such irreducible complex representations are one-dimensional, as it corresponds to the case under consideration, with symmetries that form an Abelian compact group. The disadvantage of using this complex Fourier decomposition, nonetheless, is that we have to deal with the complications posed by the reality conditions. In order to avoid these complications, and clarify how the irreducible representations over complex vector spaces combine in the case of real fields, it is then advisable to adopt an alternate decomposition in terms of real eigenfunctions of the LB operator, namely, in terms of Fourier modes corresponding to sines and cosines:
\begin{equation}\label{dec}
\varphi(t,\vec{\theta})=\frac{1}{(\pi)^{3/2}}{\sum_{\vec{n}}}'[q_{\vec{n}}(t)\cos(\vec{n}\cdot\vec{\theta}) + x_{\vec{n}}(t)\sin(\vec{n}\cdot\vec{\theta})].
\end{equation}
Here, we have defined the tuple $\vec{n}=(n_1, n_2, n_3)$, with $n_i \in \hbox{\msbm Z}$ ($i=1,2,3$). Note that we have changed the notation for the labels of the Fourier modes and coefficients from $\vec{m}$ to $\vec{n}$, to facilitate the distinction between the complex and the real formulations. The sum in the above expansion contains only {tuples $\vec{n}$ of integers whose first non-zero component is positive.} All different tuples which satisfy this restriction are to be summed over (and just once each). We indicate the restriction in the sum with a tilde, instead of making it explicit, something which would complicate the notation in excess. Furthermore, {here and} in the rest of our discussion we will ignore the zero mode, $\vec{n}=(0,0,0)$. As we can anticipate, the unitary evolution and the uniqueness of the representation do not depend on the removal of a finite number of degrees of freedom. This mode can always be quantized separately, including the possibility of employing non-standard methods in its quantum mechanical description (like, e.g., using loop quantization methods). Its exclusion does not alter the field properties of the system, on which we concentrate our attention.

Since we have expanded the field in eigenfunctions of the LB operator, the different modes are dynamically decoupled. For the cosine modes, we obtain
\begin{equation}
\ddot{q}_{\vec{n}}+[\omega^2_n + s(t)]q_{\vec{n}}=0,
\end{equation}
and similarly for the sine modes, $x_{\vec{n}}$. Thus, the dynamical equations only depend on the corresponding LB eigenvalue of the mode, $-\omega_n^2$, which is given by $\omega_n^2 = \sum_i n_i^{2}$. The subindex $n$, introduced as a label in $\omega_n$, is taken as a positive integer which designates the order of these eigenvalues. That is, $\omega_n<\omega_{n'}$ if $n<n'$. Note that, given the compactness of the three-torus, the LB eigenvalues indeed form a sequence $\{\omega_n\}$, which is certainly unbounded. It is also worth noticing that the degeneracy $g_n$ of each eigenspace of the LB operator presents a complicated dependence on the label $n$, because of accidental degeneracy: one can find different tuples $\vec{n}$ (others than those related by a flip of sign in one of the components, or by permutations of the components) which lead to the same eigenvalue. For the sake of an example: the tuples $(2,2,1)$ and $(3,0,0)$ correspond to the same eigenvalue $\omega^2=9$.

Finally, we decompose the field momentum $P_{\varphi}$ in the same way as we have explained for the field configuration. Then, the momentum coefficients for the sine and cosine contributions in the real modes expansion, which we will call $p_{\vec{n}}$ and $y_{\vec{n}}$, respectively, satisfy the dynamical equations $p_{\vec{n}}=\dot{q}_{\vec{n}}$, and $y_{\vec{n}}=\dot{x}_{\vec{n}}$. The non-vanishing Poisson brackets in terms of these coefficients are $\{q_{\vec{n}},p_{\vec{n}'}\}=\{x_{\vec{n}},y_{\vec{n}'}\}=\delta_{\vec{n}\vec{n}'}$.

\subsection{Fock quantization}
\label{s22}

The Hilbert space of the system over which we construct the quantum theory is the direct sum of the symmetric tensor products of the one-particle Hilbert space ${\cal H}_{0}$. The Fock space, then, is determined by the one-particle space. For the construction of the latter, the only ingredient needed is a complex structure $J$ \cite{wald}. We recall that a complex structure is a real map on phase space whose square is minus the identity and that preserves the symplectic structure $\Omega$ \cite{wald}. In addition, we demand the complex and the symplectic structures to be compatible in the sense that their composition, $\Omega(J\cdot,\cdot)$ must be a positive bilinear form. The sector of positive frequency (complex) solutions is obtained with the projector $\frac{1}{2} (\mathbf{1}-i J)$, where $\mathbf{1}$ is the identity. This sector is then completed into the one-particle Hilbert space using the norm provided by $\frac{1}{2} [\Omega(J\cdot,\cdot)-i\Omega(\cdot,\cdot)]$. Therefore, the complex structure contains all the information that is physically relevant to determine the different possible Fock representations. This can be rephrased by saying that a complex structure, together with the symplectic structure, defines a state, which is usually called the vacuum, and then a representation of the CCR's.

In order to discuss the Fock quantization of the scalar field, we {now} introduce a particular, well-known complex structure which we will employ as a starting point in our analysis. The chosen complex structure, $J_0$, is the one which would be naturally related to a free, massless KG field. Hence, it is determined entirely by the three-torus metric, via the corresponding LB operator. In particular, this property guarantees that the complex structure $J_0$, and thus the vacuum selected by it, is invariant under the isometry group of the three-torus. In terms of the LB operator, we can define $J_0$ using a decomposition in eigenmodes, introducing the annihilationlike variables
\begin{equation}
\label{aniquilacion}
a_{\vec{n}} = \frac{1}{\sqrt{2\omega_n}}(\omega_n q_{\vec{n}} + ip_{\vec{n}}), \qquad
\tilde{a}_{\vec{n}} = \frac{1}{\sqrt{2\omega_n}}(\omega_n x_{\vec{n}} + iy_{\vec{n}}).
\end{equation}
The corresponding creationlike variables are given by the complex conjugates $a^*_{\vec{n}}$ and $\tilde{a}^*_{\vec{n}}$. These new variables provide a complete set of coordinates on phase space. The action of $J_0$ on this new basis is defined to be diagonal, in the standard form
\begin{eqnarray}
\label{J0}
&J_0(a_{\vec{n}})=ia_{\vec{n}},\qquad &J_0(\tilde{a}_{\vec{n}})=i\tilde{a}_{\vec{n}},\\
&J_0(a^*_{\vec{n}})=-ia^*_{\vec{n}},\qquad &J_0(\tilde{a}^*_{\vec{n}})=-i\tilde{a}^*_{\vec{n}}.
\end{eqnarray}

The evolution of these variables from a fixed initial time $t_0$ to any another time $t\in\hbox{\msbm I}$ can be expressed as a linear transformation $U$ (since the field equations are linear). Given that different modes decouple in the dynamics, the evolution is actually block diagonal, with $2\times 2$ blocks $U_{\vec{n}}$, one for each pair of annihilation and creationlike variables.  Furthermore, since the dynamical equations of the modes depend only on the eigenspace of the LB operator under consideration (labeled by $n$), the same happens with the mentioned blocks, which we will therefore designate by $U_n$. Therefore, it is straightforward to conclude that the evolution can be described in the form
\begin{equation}
\label{transformacion}
\hspace*{-1.cm} \left( \begin{array}{c} a_{\vec{n}}(t) \\
a^{*}_{\vec{n}}(t)\end{array}\right) = U_n \left(
\begin{array}{c} a_{\vec{n}}(t_0) \\ a^{*}_{\vec{n}}(t_0)\end{array}\right),\quad
U_n=\left(
\begin{array}{cc} \alpha_n(t,t_0) & \beta_n(t,t_0)  \\
\beta^{*}_n(t,t_0)  & \alpha^{*}_n(t,t_0) \end{array}\right),
\end{equation}
and similarly for $(\tilde{a}_{\vec{n}},\tilde{a}^*_{\vec{n}})$. Finally, since the dynamical evolution is a symplectic transformation, the alpha and beta functions appearing in this matrix expression must satisfy the relation
\begin{equation}
|\alpha_n(t,t_0)|^{2} - |\beta_n(t,t_0)|^{2} = 1, \quad \forall t\in \hbox{\msbm I},
\end{equation}
for all values of the eigenvalue label $n$.

In the next section we will show that the Fock representation determined by $J_0$ admits a unitary implementation of the evolution and, furthermore, that this is the only representation with that property (up to unitary equivalence) among all those that are invariant under the isometry group of the three-torus. We recall that, in general, a linear canonical transformation $U$ can be implemented quantum mechanically as a unitary transformation in the representation determined by a complex structure $J$ if and only if the antilinear part of the transformation, given by $U_J=\frac{1}{2}(U+JUJ)$, is a Hilbert-Schmidt operator \cite{hr-sh}, namely, that the trace of $U_J^{\dagger}U_J$ is finite (here, the dagger denotes the adjoint operator). It is possible to rephrase this condition as the requirement that the antilinear coefficients of the considered transformation (usually called the beta Bogoliubov coefficients) be square summable (that is, that their squared norms have a convergent sum). Another equivalent way of stating this condition is to demand that the image of the vacuum state under the transformation $U$ possess a finite number of ``particles'', using the particle concept associated to the original vacuum. We will use this condition for unitary implementability (in any of its versions) in the rest of our discussion.

\section{Uniqueness of the quantization}
\label{s3}

We will now show that the Fock representation determined by the complex structure $J_0$ leads to a unitary quantum evolution, even if the field has in fact a time dependent mass. We will also characterize the most general complex structure that is invariant under the symmetry group formed by the transformations $T_{\vec{\alpha}}$. We will see that they are all related by a specific family of symplectic transformations. Using that characterization, we will prove the uniqueness of the invariant complex structure (up to unitary equivalence) under the requirement that the dynamics admit a unitary implementation. To avoid repeating parts of the demonstration which follow the line of arguments presented in the literature for the three-sphere \cite{CMV8}, we will review the main steps of the proof and concentrate our attention just on the aspects that are specific of the three-torus.

\subsection{Unitary evolution in the massless representation}
\label{s31}

According to our comments above, the evolution $U$ is implementable as a unitary transformation on the Fock space determined by $J_0$ if and only if its beta coefficients $\beta_n(t,t_0)$ [appearing in Eq. (\ref{transformacion})] are square summable. Taking into account the
degeneracy $g_n$ of the eigenspaces of the LB operator, the necessary and sufficient condition is the finiteness of the sum $\sum_n g_n |\beta_n(t,t_0)|^{2}$ for all possible values of time $t\in \hbox{\msbm I}$. Physically, this amounts to a finite particle production during
evolution. Clearly, the summability of the sequence of beta functions (in squared norm) depends on the asymptotic behavior of $\beta_n$ (and $g_n$) in the ultraviolet, namely, in the limit of infinitely large eigenvalues $\omega_n$. The asymptotic analysis of the Bogoliubov coefficients for a KG field with time dependent mass in a stationary spacetime was carried out in Ref. \cite{CMV8}, and the result applies in particular to the case discussed here, when the compact spatial sections are isomorphic to a three-torus. The analysis leads to the conclusion that
\begin{equation}\label{alpha_n}
\alpha_n(t,t_0) = e^{-i\omega_n(t-t_0)}+ O\left(\frac{1}{\omega_n}\right),\qquad
\beta_n(t,t_0) =  O\left(\frac{1}{\omega_n^2}\right).
\end{equation}
The symbol $O$ indicates the asymptotic order. The only hypothesis about the field that is employed to deduce this behavior (but that it is not even necessary for its validity) is that the mass function $s(t)$ possesses a first derivative which is integrable in every compact subinterval of $\hbox{\msbm I}$. Using the asymptotics, it is straightforward to see that the condition for a unitary implementation of the evolution is equivalent to the summability of the sequence formed by $g_n/\omega_n^4$. To check whether this summability holds, one has to study how the degeneracy $g_n$ changes with $n$ in the limit when this label gets infinitely large. This variation of $g_n$ is quite involved owing to the accidental degeneracy that we have already pointed out. The exact dependence of the degeneracy with $n$ cannot be given explicitly. Nonetheless, for our discussion, we only need to compute the asymptotic behavior of $g_n$, a task which can be done in a relatively simple way as follows.

The values $\omega_n^2$ can be understood as the norm of the vector $\vec{n}$ provided by the tuple that labels the modes. {In principle, $\vec{n}$ is restricted so that its first non-vanishing (integer) component be positive. However, there exist two modes for each value of $\vec{n}$: the sine and cosine modes. We can assign the two modes to the couple of vectors $(\vec{n},-\vec{n})$.} It is then clear that we can make correspond modes to {all} vectors with integer components (except the zero-mode, that has been excluded). Let us call $D_N$ the number of modes whose eigenvalue
$\omega_n$ is in the interval $(N,N+1]$, with $N$ a natural number. Since $1/\omega_n$ is strictly decreasing with $n$, we have that the sum $\sum_n (g_n/\omega_n^4)$, whose convergence we want to check, is always equal or smaller than $\sum_N (D_N/N^4)$. Geometrically, the value of $D_N$ is the number of vertices of the cubic lattice with step equal to one that are contained between the sphere of radius $N+1$ (including its surface) and the sphere of radius $N$. Therefore, $D_N$ increases with $N$ like $N^2$. It is then straightforward to see that the sum of $D_N/N^4$ is finite, and a fortiori that of $g_n/\omega_n^4$. Thus, the Fock representation determined by $J_0$, naturally associated with the case of a massless field, provides in fact a unitary implementation of the dynamics even in the presence of a time dependent mass.

\subsection{Characterization of the invariant complex structures}
\label{s32}

We will now characterize the most general complex structure that is invariant under the group of symmetries formed by the transformations $T_{\vec{\alpha}}$, corresponding to the isometries of the three-torus. For simplicity, we will just call {\sl invariant} such complex structures. To reach this characterization, we will follow a procedure which differs from that presented in the literature for other compact spatial topologies, adopting our analysis to the peculiarities of the three-torus. As we have remarked, in our case, the isometry group is Abelian, and hence its irreducible representations are all one-dimensional over complex vector spaces. On the other hand, since the scalar field studied is real, these representations must be combined suitably. To clarify this interplay, we will rather consider real representations from the start, adapted to the decomposition of the field in sine and cosine modes.

It is also worth noticing that, owing to the commented accidental degeneracy, the eigenspaces of the LB operator do not provide irreducible representations of the isometry group, even if the operator commutes with the isometries because it is constructed out of the three-torus metric. Again, this situation is novel compared to that found in the literature, e.g. for $d$-spheres \cite{CMV8,PRD79,CMV83,JCAP10}. Hence, the case of the three-torus calls for a detailed analysis, that we present in the rest of this subsection.

The action of the three-torus symmetries $T_{\vec{\alpha}}$ on the real sine and cosine modes [see Eq. (\ref{dec})] is easily derived from its active transformation of the field. One gets
\begin{equation}
\label{trm}
\left( \begin{array}{c} q'_{\vec{n}} \\
x'_{\vec{n}} \end{array}\right) = \left(
\begin{array}{cc} \cos(\vec{n}\cdot\vec{\alpha})  & -\sin(\vec{n}\cdot\vec{\alpha})  \\
\sin(\vec{n}\cdot\vec{\alpha})  & \;\cos(\vec{n}\cdot\vec{\alpha}) \end{array}\right) \left( \begin{array}{c} q_{\vec{n}} \\
x_{\vec{n}} \end{array}\right).
\end{equation}
Similar equations are obtained for the modes of the field momentum $(p_{\vec{n}},y_{\vec{n}})$. We notice that these transformations only mix modes in pairs, just the sine and cosine modes with the same label $\vec{n}$. Furthermore, the action on each of these pairs is different. The sine and cosine modes that get mixed belong to the same LB eigenspace and hence have the same dynamics. Besides, as we anticipated, not all modes in the same eigenspace get mixed under the action of the symmetry group, owing to the accidental degeneracy. This fact complicates the characterization of the invariant complex structures and leads to a situation which is similar to that encountered for the $S^1$ topology
\cite{ccmv1,PRD79}, though in higher dimensions. Examining the action of the transformations $T_{\vec{\alpha}}$ given above, it is not difficult to realize that the most general invariant complex structure must be block diagonal in the label $\vec{n}$, namely,
\begin{equation}
J={\bigoplus_{\vec{n}}}' J_{\vec{n}},
\end{equation}
where each complex structure $J_{\vec{n}}$ corresponds to a $4\times4$ block, associated with the sine and cosine modes determined by $\vec{n}$ for the field configuration and momentum. In the direct sum, we have used the same kind of notation introduced in Eq. (\ref{dec}).

Let us express the blocks $J_{\vec{n}}$ in terms of smaller $2\times2$ blocks in the phase (sub)space basis formed by $(q_{\vec{n}},x_{\vec{n}},p_{\vec{n}},y_{\vec{n}})$:
\begin{equation}
J_{\vec{n}}=\left(\begin{array}{cc}
A_{\vec{n}} & B_{\vec{n}}\\
C_{\vec{n}} & D_{\vec{n}}
\end{array}
\right).
\end{equation}

From the condition that $J$ be an invariant complex structure, so that $T_{\vec{\alpha}}^{-1} J T_{\vec{\alpha}}=J$ for all transformations $T_{\vec{\alpha}}$, one concludes after a straightforward computation that every $2 \times 2$ block $\{Q_{\vec{n}}\}=\{A_{\vec{n}},B_{\vec{n}},C_{\vec{n}},D_{\vec{n}}\}$ must commute with all the rotation matrices of the form
\begin{equation}
R_{\vec{n}}(\vec{\alpha})=\left(\begin{array}{cc}
\cos(\vec{n}\cdot\vec{\alpha}) & - \sin(\vec{n}\cdot\vec{\alpha})\\
\sin(\vec{n}\cdot\vec{\alpha}) & \;\cos(\vec{n}\cdot\vec{\alpha})
\end{array}
\right),
\end{equation}
that is
\begin{equation}
R^{-1}_{\vec{n}}(\vec{\alpha})Q_{\vec{n}}R_{\vec{n}}(\vec{\alpha})=Q_{\vec{n}}.
\end{equation}
It then follows that every $2\times2$ block must have a diagonal part proportional to the identity and a skew-symmetric non-diagonal part, namely,
\begin{equation}
Q_{\vec{n}}=\left(\begin{array}{cc}
\;Q_{\vec{n}}^{(1)} & Q_{\vec{n}}^{(2)}\\
-Q_{\vec{n}}^{(2)} & Q_{\vec{n}}^{(1)}
\end{array}
\right),
\end{equation}
where $Q_{\vec{n}}^{(1)}$ and $Q_{\vec{n}}^{(2)}$ are arbitrary real numbers.

We still have to impose the condition that the invariant complex structure be compatible with the symplectic structure, so that their combination $\Omega(J\cdot,\cdot)$ provides a positive definite bilinear map on phase space. In terms of the blocks $J_{\vec{n}}$, this condition implies that $J^{T}_{\vec{n}}\Omega_{\vec{n}}$ must be a positive definite symmetric matrix, where $J^{T}_{\vec{n}}$ is the transpose of $J_{\vec{n}}$ and the blocks of the symplectic structure are
\begin{equation}
\Omega_{\vec{n}}=\left(\begin{array}{cc} \mathbf{0}_{2\times2} & -\mathbf{1}_{2\times2}\\ \mathbf{1}_{2\times2} & \,\,\,\mathbf{0}_{2\times2}
\end{array}
\right).
\end{equation}
Here, $\mathbf{0}_{2\times2}$ is the zero matrix in two dimensions and $\mathbf{1}_{2\times2}$ is the identity matrix. In order to satisfy this requirement, the $2\times 2$ blocks of $J_{\vec{n}}$ given by $B_{\vec{n}}$ and $C_{\vec{n}}$ must be symmetric matrices of negative and positive definite type, respectively, whereas the two other blocks must satisfy that $A_{\vec{n}}=-D_{\vec{n}}$. Together with the condition of invariance discussed above, we then conclude that the blocks $J_{\vec{n}}$ must have the form
\begin{equation}
J_{\vec{n}}=\left(\begin{array}{cc} A_{\vec{n}} & \; B_{\vec{n}}\\ C_{\vec{n}} & -A_{\vec{n}}
\end{array}
\right),
\end{equation}
where $B_{\vec{n}}$ and $C_{\vec{n}}$ must be proportional to the identity, with a non-positive and a non-negative constant of proportionality, respectively. That is,
\begin{equation}
B_{\vec{n}}^{(2)}=C_{\vec{n}}^{(2)}=0,\qquad B_{\vec{n}}^{(1)}\le 0,\qquad C_{\vec{n}}^{(1)}\ge 0.
\end{equation}

In addition, we must also impose the remaining conditions that the square of the complex structure be minus the identity, $J^{2}=-\mathbf{1}$, and that it leave invariant the symplectic structure, namely $\Omega(J\cdot,J\cdot)=\Omega(\cdot,\cdot)$. These requirements lead in turn to the following conditions on the blocks $J_{\vec{n}}$:
\begin{equation}
J^{2}_{\vec{n}}=-\mathbf{1}_{4\times 4},\qquad
J^{T}_{\vec{n}}\Omega_{\vec{n}}J_{\vec{n}}=\Omega_{\vec{n}}.
\end{equation}
From the former of these restrictions we obtain the following equations for the matrix elements:
\begin{equation}
\label{jjeq}
-\big[A_{\vec{n}}^{(1)}\big]^{2}+\big[A_{\vec{n}}^{(2)}\big]^{2}-B_{\vec{n}}^{(1)}C_{\vec{n}}^{(1)}=1,
\qquad A_{\vec{n}}^{(1)}A_{\vec{n}}^{(2)}=0.
\end{equation}
On the other hand, the second restriction leads to the equations
\begin{eqnarray}
\label{jojeq}
&&\big[A_{\vec{n}}^{(1)}\big]^{2}+\big[A_{\vec{n}}^{(2)}\big]^{2}+B_{\vec{n}}^{(1)}C_{\vec{n}}^{(1)}=-1, \\
&& A_{\vec{n}}^{(2)}B_{\vec{n}}^{(1)}=0, \qquad A_{\vec{n}}^{(2)}C_{\vec{n}}^{(1)}=0.
\end{eqnarray}
Summing the first equality of Eq. (\ref{jjeq}) and Eq. (\ref{jojeq}), we get that the matrix element $A_{\vec{n}}^{(2)}$ mush vanish. Therefore, every block $J_{\vec{n}}$ of an invariant complex structure (compatible with the symplectic form) must have the form
\begin{equation}
J_{\vec{n}}=\left(\begin{array}{cc} A_{\vec{n}}^{(1)}\mathbf{1}_{2\times2} & \; B_{\vec{n}}^{(1)}\mathbf{1}_{2\times2}\\ C_{\vec{n}}^{(1)}\mathbf{1}_{2\times2} & -A_{\vec{n}}^{(1)}\mathbf{1}_{2\times2}
\end{array}
\right),
\end{equation}
where
\begin{equation}
B_{\vec{n}}^{(1)}<0,\qquad C_{\vec{n}}^{(1)}>0,\qquad A_{\vec{n}}^{(1)}=\pm\sqrt{-1-B_{\vec{n}}^{(1)}C_{\vec{n}}^{(1)}}.
\end{equation}
Note that $B_{\vec{n}}^{(1)}C_{\vec{n}}^{(1)}\le -1$, since $A_{\vec{n}}^{(1)}$ must be a real number.

Once we have deduced the general expression of the invariant complex structures in the basis $\{(q_{\vec{n}},x_{\vec{n}},p_{\vec{n}},y_{\vec{n}})\}$, it is straightforward to write them in the basis $\{(a_{\vec{n}},a^{\ast}_{\vec{n}},\tilde{a}_{\vec{n}},\tilde{a}^{\ast}_{\vec{n}})\}$ of annihilation and creationlike variables defined by the complex structure $J_{0}$, basis in which we can easily compare them. The considered change of basis can be obtained from Eqs. (\ref{aniquilacion}). In this way, one can check that the blocks $J_{\vec{n}}$ of any admissible invariant complex structure must be block diagonal, with $2\times2$ blocks that coincide by pairs. More explicitly,
\begin{equation}
J_{\vec{n}}=\left(\begin{array}{cc}
\mathcal{J}_{\vec{n}} & \mathbf{0}_{2\times2}\\ \;\mathbf{0}_{2\times2} &  \mathcal{J}_{\vec{n}}
\end{array} \right), \end{equation}
with
\begin{equation}
\mathcal{J}_{\vec{n}}= i\left(\begin{array}{cc}
-B_{\vec{n}}^{(1)} + C_{\vec{n}}^{(1)}\omega^{2}_{n} &
-B_{\vec{n}}^{(1)} - C_{\vec{n}}^{(1)}\omega^{2}_{n}+ i A_{\vec{n}}^{(1)}\omega_{n}\\
B_{\vec{n}}^{(1)}+C_{\vec{n}}^{(1)}\omega^{2}_{n}+ i
A_{\vec{n}}^{(1)}\omega_{n} &
B_{\vec{n}}^{(1)} - C_{\vec{n}}^{(1)}\omega^{2}_{n}
\end{array}
\right).
\end{equation}

Actually, every invariant complex structure $J$ (that is compatible with the symplectic structure) can be obtained from $J_0$ by means of a symplectic transformation, $K$, namely $J=KJ_0K^{-1}$ \cite{ccmv1}. Since, in the considered basis of annihilation and creationlike variables, $J_0$ is diagonal with blocks of the form $(J_0)_{\vec{n}}={\rm diag}\{i,-i,i,-i\}$, the transformation $K$ can also be taken block diagonal, with $4\times4$ blocks of the type
\begin{equation}
\label{st1}
K_{\vec{n}}=\left(\begin{array}{cc}
\mathcal{K}_{\vec{n}} & \mathbf{0}_{2\times2} \\ \mathbf{0}_{2\times2} & \mathcal{K}_{\vec{n}}
\end{array}\right), \qquad
\mathcal{K}_{\vec{n}} = \left(\begin{array}{cc}
\kappa_{\vec{n}} & \lambda_{\vec{n}} \\
\lambda_{\vec{n}}^{\ast} & \kappa_{\vec{n}}^{\ast}
\end{array}\right).
\end{equation}
Here, $\kappa_{\vec{n}}$ and $\lambda_{\vec{n}}$ are complex numbers which play the role of alpha and beta Bogoliubov coefficients for the transformation $K$. In particular, they satisfy the symplectomorphism condition
\begin{equation}\label{Kcoeff}
|\kappa_{\vec{n}}|^{2}-|\lambda_{\vec{n}}|^{2}=1\qquad \forall \vec{n}.
\end{equation}
Finally, the relation of the coefficients $\kappa_{\vec{n}}$ and $\lambda_{\vec{n}}$ with the matrix elements $A_{\vec{n}}^{(1)}$, $B_{\vec{n}}^{(1)}$, and $C_{\vec{n}}^{(1)}$ is given by
\begin{eqnarray}
&&2|\kappa_{\vec{n}}|^2=1-B_{\vec{n}}^{(1)} + C_{\vec{n}}^{(1)}\omega^{2}_{n}, \\
&& 2  \kappa_{\vec{n}}\lambda_{\vec{n}}=B_{\vec{n}}^{(1)}+C_{\vec{n}}^{(1)}\omega^{2}_{n}- i A_{\vec{n}}^{(1)}\omega_{n}.
\end{eqnarray}
Notice that the phase of $\kappa_{\vec{n}}$ can be chosen freely. For instance one can choose it so that this Bogoliubov coefficient be non-negative. Actually, one can see that this choice does not affect the rest of our considerations.

\subsection{Uniqueness of the invariant representation with unitary dynamics}
\label{s33}

To conclude our proof, showing the validity of our criterion to pick out a unique Fock representation, we still have to consider the unitary implementation of the dynamics and demonstrate that this restricts the admissible invariant complex structures to only one class of unitary equivalence. We will use again the fact that all the possible invariant complex structures $J$ are related with the complex
structure  $J_0$ by means of a symplectomorphism $K$ of the form given above, which can be understood just as a change of annihilation and creationlike variables. It follows that the unitary implementation of the evolution $U$ in the representation determined by $J$ amounts to the unitary
implementation  of $K^{-1}UK$ with respect to $J_0$ \cite{ccmv1}. The beta coefficients of $K^{-1}UK$ can be viewed as the antilinear coefficients of the Bogoliubov transformation determined by the dynamics, expressed in terms of the annihilation and creationlike operators selected by $J$, instead of by $J_0$. A trivial computation shows that these new beta coefficients, that we will call $\beta_{\vec{n}}^{J}$, take the following expression in terms of the original Bogoliubov coefficients:
\begin{equation}
\label{nbac1}
\beta_{\vec{n}}^{J}(t,t_0)=(\kappa_{\vec{n}}^{\ast})^{2}\beta_{n}(t,t_0)-
(\lambda_{\vec{n}})^{2}\beta_{n}^{\ast}(t,t_0)+2i\kappa_{\vec{n}}^{\ast}\lambda_{\vec{n}}\hbox{\frakfamily I}[\alpha_{n}(t,t_0)].
\end{equation}
Here, $\hbox{\frakfamily I}[\cdot]$ denotes the imaginary part. It is worth remarking that these beta coefficients (or rather beta functions, taking into account the time variation) depend now not just on $n$, the label of the eigenspaces of the LB operator, but rather on $\vec{n}$, which is the label of the sine and cosine modes and, as a consequence, of the ``irreducible'' {\sl real} representations of the symmetry group of the three-torus. Therefore, if, according to our criterion, we restrict our discussion exclusively to invariant complex structures $J$ which allow a unitary quantum evolution, the above beta functions will have to be square summable (over $\vec{n}$) at all values of time $t\in\hbox{\msbm I}$. So, we will assume that this is the case from now on.

The analysis of this summability can be made along a line of arguments similar to that presented in Ref. \cite{CMV8}. Using that the beta functions $\beta_{n}(t,t_0)$ corresponding to $J_0$ are square summable and that $|\kappa_{\vec{n}}|\ge1$,
one can prove that the summability (in square norm) of $\beta_{\vec{n}}^{J}(t,t_0)$ implies the same property for the set formed by $\hbox{\frakfamily I}[\alpha_{n}(t,t_{0})]\lambda_{\vec{n}}/\kappa_{\vec{n}}^{\ast}$. Recalling then the asymptotic behavior of $\alpha_{n}(t,t_0)$ and calling $z_{\vec{n}}=\lambda_{\vec{n}}/\kappa_{\vec{n}}^{\ast}$, we arrive at the conclusion that the quantities
\begin{equation}
\label{ssc1}
z_{\vec{n}}\sin\left[\omega_{n}(t-t_0)+\frac{1}{2\omega_{n}}\int_{t_{0}}^{t}d\bar{t}\,s(\bar{t})\right]
\end{equation}
form a set which is square summable. The deduction of this result assumes, as a sufficient (but not necessary) condition, that the mass function $s(t)$ possesses a second derivative which is integrable in every compact subinterval of the time domain $\hbox{\msbm I}$. A time integration, over any such subinterval, of the partial sums of the square norms of the elements (\ref{ssc1}), combined with a suitable application of Luzin's theorem \cite{Luzin} (which is possible because the considered elements are measurable functions), shows then that the set formed by $z_{\vec{n}}$ (namely, the ratios of the coefficients of $\mathcal{K}_{\vec{n}}$) is square summable. Given the definition of $z_{\vec{n}}$ and relation (\ref{Kcoeff}), one can see that this implies that the set formed by the antilinear coefficients of $K$, $\lambda_{\vec{n}}$, is square summable as well \cite{CMV8}. But this last summability result is precisely the condition for the unitary implementability of the symplectomorphism $K$ and, hence, of the unitary equivalence of the two Fock representations determined the complex structures $J$ and $J_0$, related by that symplectomorphism. Since the discussion is valid for all admissible invariant complex structures $J$, we conclude that all such structures which besides allow for a unitary dynamics are indeed equivalent. Thus, our criterion of invariance under the three-torus isometries and of unitarity in the evolution picks out a unique {equivalence class of (invariant) representations} for the scalar field.

\section{Uniqueness of the field description}
\label{s4}

We will show now that our criterion is not only capable of selecting a unique equivalence class of Fock representations for the quantization of a KG field with a time dependent mass in a flat spacetime with the spatial topology of a three-torus, but, beyond that, it also determines a unique canonical pair for the field among all those that are related by a linear canonical transformation varying in time and in which the field configuration gets scaled. This kind of scaling transformations are often found in cosmological contexts, either when dealing with test fields or with perturbations around homogeneous and isotropic solutions, which can represent a genuine background spacetime, an effective spacetime on which the propagation takes place once certain quantum effects are taken into account, or just an auxiliary background in which one formulates in a simpler form the field dynamics (for instance after dimensional reduction in symmetric models in General Relativity). The scaling absorbs part of the time dependence of the field, which is assigned to the time variation of the background. Although the classical formulations obtained for the field with this class of transformations are all equivalent, this ceases to be the case in the quantum theory, both because not all linear canonical transformations admit a unitary implementation and because the time dependence of the transformation changes the dynamics, in particular affecting its properties of unitarity. As a consequence, the criterion of a unitary evolution has different implications in the distinct formulations reached with these transformations.

\subsection{Time dependent canonical transformations}
\label{s41}

The most general linear canonical transformation which includes a scaling of the field configuration and allows for a time dependence of the linear coefficients has the form
\begin{equation}
\label{tds1}
\phi=f(t)\varphi, \qquad P_{\phi}=\frac{P_{\varphi}}{f(t)}+g(t)\varphi.
\end{equation}
In the last equation, we have taken again into account that the determinant of the three-torus metric is equal to one. We assume that the real functions $f(t)$ and $g(t)$ that characterize the transformation are at least twice differentiable, in order to respect the differentiability properties in the field equations. Besides, we suppose that the function $f(t)$ never vanishes, so that no spurious singularity is introduced in the field with the considered transformation. Finally, by means of a constant canonical linear transformation (which does not change the Fock representation of the field system), we can always set the initial values of the two time functions involved in our change equal to $f(t_0)=1$ and $g(t_0)=0$ \cite{PRD75}, so that the original and the transformed canonical pairs coincide initially.

For the new pair $(\phi,P_{\phi})$, we start adopting the representation determined by the complex structure $J_0$. The time dependent transformation which leads to this pair changes the dynamics with respect to the original one, $U$. The new dynamical transformation $\tilde{U}$ has $2\times 2$ blocks labeled again by the LB eigenvalue number $n$, and given by \cite{CMV83}:
\begin{equation} \label{3} \tilde{U}_n(t,t_0)= C_n(t) U_n(t,t_0),
\end{equation}
where
\begin{eqnarray} \label{2} C_n(t) &:=&
\left(\begin{array}{cc} f_{+}(t)+iG_n(t) \;&
f_{-}(t)+iG_n(t) \\ f_{-}(t)-
iG_n(t)\;& f_{+}(t)-iG_n(t)
\end{array} \right),\\
2 f_{\pm}(t)&:=&f(t)\pm \frac{1}{f(t)},\qquad G_n(t)=\frac{g(t)}{2\omega_n}.
\end{eqnarray}

On the other hand, let us recall that the most general invariant complex structure $J$, compatible with the symplectic structure, has already been characterized  in Sec. \ref{s32}: it is related with $J_0$ by a symplectomorphism $K$ of the type (\ref{st1}). In total, then, we see that the new dynamics $\tilde{U}$ turns out to admit a unitary implementation with respect to an invariant complex structure $J$ if and only if the beta Bogoliubov functions of the transformations with blocks $\mathcal{K}_{\vec{n}}^{-1}C_n(t) U_n(t,t_0)\mathcal{K}_{\vec{n}}$ are square summable over all the possible values of $\vec{n}$ (the label of the sine and cosine modes). These beta functions, that we will call $\tilde{\beta}^{J}_{\vec{n}}(t,t_{0})$, adopt an expression similar to that given in Eq. (\ref{nbac1}), but with the Bogoliubov functions $\alpha_n(t,t_0)$ and $\beta_n(t,t_0)$ --corresponding to the reference complex structure $J_0$-- replaced with those of the evolution $\tilde{U}$ for the pair $(\phi,P_{\phi})$ (see Ref. \cite{CMV83}):
\begin{equation}
\label{nabf1}
\tilde{\alpha}_{n}(t,t_0)=f_{+}(t)\alpha_{n}(t,t_0)+f_{-}(t)\beta_{n}^{\ast}(t,t_0)
+iG_n(t)[\alpha_{n}(t,t_{0})+\beta_{n}^{\ast}(t,t_0)],
\end{equation}
\begin{equation}
\label{nabf2}
\tilde{\beta}_{n}(t,t_0)=f_{+}(t)\beta_{n}(t,t_0)+f_{-}(t)\alpha_{n}^{\ast}(t,t_0)
+iG_n(t)[\beta_{n}(t,t_{0})+\alpha_{n}^{\ast}(t,t_0)].
\end{equation}
In the rest of this section, we will demonstrate that a unitary dynamics is possible only if $f(t)=1$ and $g(t)=0$ at all times, that is, if we describe our field precisely with the original canonical pair, associated with the formulation as a KG field in flat spacetime with compact spatial sections and a time dependent mass.

\subsection{Uniqueness of the scaling}
\label{s42}

We will first show that $f(t)$ must be the identity function in $\hbox{\msbm I}$. In order to do this, we simply adapt the proof explained in
{Ref. \cite{gr-qc}}. For each eigenvalue of the LB operator, $-\omega_n^2$, let us choose a value $\vec{M}_n$ of the label $\vec{n}$ among all those whose Euclidean norm as a vector coincides with $\omega_n$. We then consider the sequence with elements $\tilde{\beta}^{J}_{\vec{M}_n}(t,t_0)$. This sequence is a subset of the beta functions $\tilde{\beta}^{J}_{\vec{n}}(t,t_0)$, obtained by ignoring the degeneracy of the LB eigenspaces. Since the latter set is square summable if we admit that the dynamics is unitary, then, {a fortiori}, the sequence with labels $\vec{M}_n$ is also square summable at all times. Recalling the asymptotic behavior (\ref{alpha_n}) and employing that $|\kappa_{\vec{M}_n}|\ge 1$, it is not difficult to check that the considered square summability implies a vanishing limit at all times for the sequence with terms
\begin{equation}
\label{z1}
\left[e^{i\omega_n(t-t_0)}-z^2_{\vec{M}_{n}}e^{-i\omega_n(t-t_0)}\right]f_-(t)-2iz_{\vec{M}_{n}} \sin[\omega_{n}(t-t_0)] f_{+}(t).
\end{equation}
Let us introduce {now} the real and imaginary parts of  $z_{\vec{M}_{n}}$:
\begin{equation}
z_{\vec{M}_{n}}=\Re_{\vec{M}_{n}}+i\Im_{\vec{M}_{n}}.
\end{equation}
A straightforward computation shows that a necessary condition for the {vanishing limit of the (complex) sequence (\ref{z1})
is that the sequence}
\begin{equation}
\label{master}
 f_{-}\,(\Re^2_{\vec{M}_{n}}+\Im^2_{\vec{M}_{n}}-1)[(1+\Re^2_{\vec{M}_{n}} +\Im^2_{\vec{M}_{n}})f_{-}-2\Re_{\vec{M}_{n}}f_{+}]
\end{equation}
{vanish as well in} the limit $n\to \infty$, at all  times {[to} simplify the notation, we have {obviated} the explicit time dependence of
the functions $f_{\pm}(t)$]. {On the other hand, detailed arguments developed in Ref. \cite{gr-qc} demonstrate that} a further necessary condition {for} the unitary implementability of the dynamics is that the sequence of elements $(\Re^2_{\vec{M}_{n}}+\Im^2_{\vec{M}_{n}}-1)$ does not tend to zero.

{The last step in our proof is to show that, then, the unitary implementation is not possible unless the function $f(t)$ is the unit function. Let us suppose just the opposite, namely, that $f(t)$ is not identically the unit function. Hence, at certain values of the time $t$, we will have that $f(t)\not =1$. We will focus our discussion on those values of $t$ and see that we arrive in fact at a contradiction. Notice that in the points that we are considering, we get $f_{-}(t)\not =0$. Besides, recall that $f(t)$ is a positive and continuous function (actually, we have assumed that it is twice
differentiable).}

{Returning to expression
(\ref{master}), a necessary condition for the unitary implementation of the
dynamics is that the sequence with elements
\begin{equation}
\label{smaster}
(\Re^2_{\vec{M}_{n}}+\Im^2_{\vec{M}_{n}}-1)[(1+\Re^2_{\vec{M}_{n}} +\Im^2_{\vec{M}_{n}})f_{-}-2\Re_{\vec{M}_{n}}f_{+}]
\end{equation}
{tends to zero at all the instants of $t$ under consideration. In addition, since we know that the
sequence formed by $(\Re^2_{\vec{M}_{n}}+\Im^2_{\vec{M}_{n}}-1)$ cannot tend to zero at
large $n$ \cite{gr-qc}, we can assure that there exists a number $\epsilon>0$ and a subsequence $S$ of positive
integers $n$ such that $|\Re^2_{\vec{M}_{n}}+\Im^2_{\vec{M}_{n}}-1|>\epsilon$ in $S$. Clearly, this fact
implies that the second factor that appears in Eq. (\ref{smaster}) must have a vanishing limit on that
subsequence. Using this result we straightforwardly deduce that the following expression must have a zero limit on the studied subsequence $S$:}
\begin{equation}
\label{firstproof}
f^2(t)[(1- \Re_{\vec{M}_{n}})^2+\Im^2_{\vec{M}_{n}}] - [(1+ \Re_{\vec{M}_{n}})^2+\Im^2_{\vec{M}_{n}}].
\end{equation}
But, given that the two time independent sequences $(1- \Re_{\vec{M}_{n}})^2+\Im^2_{\vec{M}_{n}}$ and
$(1+ \Re_{\vec{M}_{n}})^2+\Im^2_{\vec{M}_{n}}$ cannot both tend to zero, the vanishing of the limit of the above expression immediately requires that the function $f(t)$ take exactly the same value at at all the instants of time that we are considering (namely, those where $f(t) \not = 1$). In this way, we reach the conclusion that the function $f(t)$ can take at most two distinct values: one of them equal to 1 (e.g., at the reference time $t_0$) and maybe another value which has been assumed to be different from the unity. However, such a behavior is precluded by the continuity of the function. This clear contradiction proves that the only consistent possibility is that $f(t)$ is indeed identically equal to the unit function, as we wanted to demonstrate.}

In this way, we reach the result that no scaling of the field configuration is allowed by our combined criterion of invariance under the three-torus isometries and the unitarity of the dynamics.

\subsection{Uniqueness of the field momentum}
\label{s43}

We will end this section by proving that no change in the momentum is permitted by the requirement of symmetry invariance and unitary evolution, so that the function $g(t)$ must vanish. We return to the expression of the beta functions for the dynamics of the system after performing a time dependent linear canonical transformation, but now specialized to the case $f(t)=1$, in accordance with the discussion of the previous subsection. The demand that, for an invariant complex structure $J$, the dynamics admit a unitary implementation amounts to the square summability of the set formed by $\tilde{\beta}^{J}_{\vec{n}}(t,t_0)$ at all instants of time. Recalling that $|\kappa_{\vec{n}}|\ge 1$, this summability ensures the same property for the set given by $\tilde{\beta}^{J}_{\vec{n}}(t,t_0)/(\kappa_{\vec{n}}^{\ast})^{2}$. Then, using the asymptotic relations (\ref{alpha_n}), and that $|z_{\vec{n}}|\le1$, one can deduce that the set formed by
\begin{eqnarray}\label{fin}
&&
G_n(t)\left\{e^{i[\omega_{n}(t-t_0)-\delta_{\vec{n}}]}+|z_{\vec{n}}|^{2}e^{-i[\omega_{n}(t-t_0)-\delta_{\vec{n}}]}+2|z_{\vec{n}}|
\cos[\omega_{n}(t-t_0)]\right\}\nonumber\\ &&
+2|z_{\vec{n}}|\hbox{\frakfamily I}\left[\alpha_{n}(t,t_0)\right]
\end{eqnarray}
is square summable at all times in the considered interval $\hbox{\msbm I}$. Here, $\delta_{\vec{n}}$ is the phase of $z_{\vec{n}}$. Obviously, the square summability is also true for the set obtained by dividing those terms by $\omega_n$, since this eigenvalue provides a sequence that diverges to infinity. Hence, employing that $g_{n}/\omega^{4}_{n}$ is a summable sequence (as we proved in Sec. \ref{s31}) and the definition of $G_n(t)$, we conclude that the set with elements
$|z_{\vec{n}}|\hbox{\frakfamily I}\left[\alpha_{n}(t,t_0)\right]/\omega_{n}$ must also be square summable. By performing a convenient time average and making use of Luzin's theorem, one can show then that the set formed by $|z_{\vec{n}}|/\omega_{n}$ has to be square summable as well.

Taking into account this result in the consideration of the terms (\ref{fin}), we arrive at the square summability of the set formed by
\begin{equation}
G_n(t) e^{ i[\omega_{n}(t-t_0)-\delta_{\vec{n}}]}+2|z_{\vec{n}}|\hbox{\frakfamily I}\left[\alpha_{n}(t,t_0)\right].
\end{equation}
In particular, the imaginary part of these quantities, namely
\begin{equation}
\frac{g(t)}{2\omega_n}\sin{[\omega_{n}(t-t_0)-\delta_{\vec{n}}]}\end{equation}
[where we have used again the definition of $G_n(t)$], is also a square summable set at all times. If there existed a subinterval of $\hbox{\msbm I}$ where the function $g(t)$ did not vanish, a suitable time integration over it would lead to the conclusion that the sequence with elements $g_{n}/\omega^{2}_n$ must be summable \cite{gr-qc}. But this sequence has in fact a divergent sum, because the sum exceeds that of $D_{N}/(N+1)^2$ over the positive integers, which clearly diverges, given the asymptotic behavior $D_{N} \propto N^{2}$ discussed in Sec. \ref{s31}. This eliminates the possibility that $g(t)$ may differ from zero in any subinterval of $\hbox{\msbm I}$. Since the function is continuous, this implies that $g(t)$ has to be the zero function.

Summarizing, our criterion determines completely the choice of canonical pair for the field among all the possibilities related by means of a time dependent linear canonical transformation. The criterion removes the freedom to scale the field, and to redefine its momentum by including a contribution that is linear in the field configuration.

\section{Conclusions}
\label{s5}

We have analyzed two types of ambiguities that appear in the Fock quantization of scalar fields in cosmological spacetimes with spatial sections that are isomorphic to a three-torus. The first of these ambiguities is related to the possibility of scaling the field by means of a time dependent function, which assigns to the (physical, effective or auxiliary) background cosmological spacetime part of the time variation. This scaling can be viewed as resulting from a linear canonical transformation, in which the field momentum gets the inverse scaling (compared to the field configuration). In addition, the momentum may admit a linear contribution of the field configuration, with a time dependent coefficient. Canonical transformations of this kind often lead to a simpler and better behaved formulation for the system \cite{mukhanov}. Each of these transformations provides a different canonical pair for the field description and, furthermore, changes the dynamics, since part of the time dependence is absorbed in the background. The other ambiguity that we have considered refers to the possible choices of Fock representation for the CCR's, once a specific canonical pair (and dynamical evolution) is given for the field. The physically different representations can be understood as corresponding to inequivalent choices of vacuum for the Fock construction. Alternatively, the physical freedom in the selection of a representation can be assigned to the possible inequivalent choices of complex structure. This ambiguity is well known in quantum field theory \cite{wald}, and it is common to remove it by introducing certain requirements on the vacuum, or equivalently on the complex structure, such as incorporating certain symmetries of the background spacetime, or presenting a especially good local or dynamical behavior.

We have put forward a criterion to remove these ambiguities in systems that, by means of one of the considered time dependent canonical transformations, can be formulated as a KG field with time varying mass propagating in a flat spacetime with (compact) three-torus spatial topology. The criterion consists of two requirements. First, that the complex structure (and hence the vacuum) be invariant under the isometries of the three-torus, equipped with the standard metric. Second, that this complex structure also allows for the unitary implementation of the dynamics. Notice, in particular, that the considered ambiguity under time dependent canonical transformations affects the dynamical evolution of the system, so that our demand of unitarity has different implications for the distinct field descriptions obtained with such transformations. On the other hand, requiring a unitary dynamics guarantees that the standard probabilistic interpretation of quantum field theory is consistent in the evolution, so that one does not have to renounce to it.

We have demonstrated that our criterion is indeed capable of removing the two mentioned types of ambiguities, selecting a unique canonical pair for the field among all those related by linear canonical transformations with time dependent coefficients and, furthermore, selecting a unique class of unitarily equivalent Fock representations for the corresponding CCR's. This class contains the representation that would be naturally associated with the case of a massless field, though now employed to define a Fock quantization of a system that is not only massive, but whose mass changes in time. To obtain this uniqueness result, we have assumed only a very mild restriction on the time dependent mass: that it must possess a second derivative which is integrable in every compact subinterval of the time domain. In fact, this assumption is just a sufficient condition, but does not even seem to be strictly a necessary one. Let us also emphasize that the conclusion about the uniqueness of the Fock quantization is valid for {\sl any} possible time domain provided that it is a (non-infinitesimal) interval of the real line.

Our result about the uniqueness of the Fock quantization can be applied to a number of physically interesting situations in cosmology. Current observations indicate that the large scale structure of the universe is (approximately) homogeneous and isotropic. Therefore, (quantum) matter fields in cosmology are naturally described by quantum field theory in FRW spacetimes. Besides, inhomogeneities can then be treated as perturbations in a fairly good approximation \cite{mukhanov2}. At leading order, these perturbations are also described as linear fields that propagate in FRW spacetimes. Furthermore, the observations favor a spatially flat cosmology, which is precisely the case studied in this work. On the other hand, the fact that the spatial topology is assumed to be compact should not pose a severe restriction, because one would expect that, beyond a certain cosmological scale related with the Hubble radius, physical interactions should have no relevant effect. Then, the physics would not be altered importantly by considering a compactification scale larger than this cosmological one. In this general context, a simple system that can be reformulated by means of a scaling as a KG field with time dependent mass in a flat spacetime is a massive, minimally coupled scalar field. A description of this type is also obtained after a suitable scaling for cosmological perturbations in conformal time. For instance, this is the case of tensor perturbations (for which one can generalize our considerations, presented here for scalar fields) and of the gauge invariant energy density perturbation amplitude \cite{mukhanov,bardeen}. In addition, one obtains a similar kind of field description for scalar perturbations of a massive scalar field around flat FRW spacetimes after adopting (e.g.) a longitudinal gauge, with the only caveat that the KG equation is modified with subdominant terms which, nonetheless, do not affect the asymptotic behavior employed in our discussion \cite{PRD85}.\footnote{Actually, the unique Fock quantization selected by our criterion for this gauge fixed system is unitarily equivalent to the quantization picked out for the gauge invariant energy density perturbation \cite{PRD85}, reassuring the consistency of our approach.} The fact that, in all these cases, the criterion of spatial symmetry invariance and unitary evolution turns out to determine a unique Fock quantum description provides the quantum field theory and its predictions with the desired robustness.

\section*{Acknowledgements}

We acknowledge financial aid from the research grants MICINN/MINECO FIS2011-30145-C03-02 from Spain,
CERN/FP/116373/2010 from Portugal, DGAPA-UNAM IN117012-3 from Mexico, and CONICYT/FONDECYT\-/\-POSTDOCTORADO/3140409 from Chile. {J.C. is grateful to the Department of Physics at UAM-I for hospitality during sabbatical leave.}


\begin{thebibliography}{99}
\bibitem{Vonneu} B. Simon, {\it Topics in Functional Analysis}, edited by R. F. Streater (Academic Press, London, England, 1972).

\bibitem{wald} R. M. Wald, {\it Quantum Field Theory in Curved Spacetime and Black Hole Thermodynamics} (Chicago University Press, Chicago, U.S.A., 1994).

\bibitem{a-m} A. Ashtekar and A. Magnon, Proc. R. Soc. Lond. {\bf A 346}, 375 (1975); A. Ashtekar and A. Magnon-Ashtekar, Pramana {\bf 15}, 107 (1980).

\bibitem{jackiw} R. Floreanini, C. T. Hill, and R. Jackiw, Ann. Phys. (New York) {\bf 175}, 345 (1987).

\bibitem{mukhanov} V. Mukhanov, {\it Physical Foundations of Cosmology} (Cambridge University Press, Cambridge, England, 2005).

\bibitem{b-d} N. D. Birrell and P. C. W. Davies, {\it Quantum Fields in Curved Space} (Cambridge University Press, Cambridge, England, 1982).

\bibitem{AKL} A. Ashtekar, W. Kaminski, and J. Lewandowski, Phys. Rev. D {\bf 79}, 064030 (2009).

\bibitem{PRD85} M. Fern\'andez-M\'endez, G. A. Mena Marug\'an, J. Olmedo, and J. M. Velhinho, Phys. Rev. D {\bf 85}, 103525 (2012).

\bibitem{PRD86} M. Fern\'andez-M\'endez, G. A. Mena Marug\'an, and J. Olmedo, Phys. Rev. D {\bf 86}, 024003 (2012).

\bibitem{PRD88} {M. Fern\'andez-M\'endez, G.A. Mena Marug\'an, and J. Olmedo, Phys. Rev. D {\bf 88}, 044013 (2013).}

\bibitem{Gowdy} R. H. Gowdy, Ann. Phys. (New York) {\bf 83}, 203 (1974).

\bibitem{PRD72} J. Cortez and G. A. Mena Marug\'an, Phys. Rev. D {\bf 72}, 064020 (2005).

\bibitem{unit-gt3} A. Corichi, J. Cortez, and G. A. Mena Marug\'an, Phys. Rev. D {\bf 73}, 041502 (2006); Phys. Rev. D {\bf 73}, 084020 (2006).

\bibitem{muksasaki} V. F. Mukhanov, JETP ett. {\bf 41}, 493 (1985); M. Sasaki, Prog. Theor. Phys. {\bf 76}, 1036 (1986).

\bibitem{generalization} {L. Castell\'o Gomar and G. A. Mena Marug\'an, {\it Uniqueness of the Fock Quantization of Scalar Fields and Processes with Signature Change in Cosmology}, arXiv:1403.6984, Phys. Rev. D in press (2014).}

\bibitem{ccmv1} A. Corichi, J. Cortez, G. A. Mena Marug\'an, and J. M. Velhinho, Class. Quantum Grav. {\bf 23}, 6301 (2006).

\bibitem{PRD76} A. Corichi, J. Cortez, G. A. Mena Marug\'an, and J. M. Velhinho, Phys. Rev. D {\bf 76}, 124031 (2007).

\bibitem{PRD75} J. Cortez, G. A. Mena Marug\'an, and J. M. Velhinho, Phys. Rev. D {\bf 75}, 084027 (2007).

\bibitem{BVV2} J. F. Barbero G., D. G. Vergel, and E. J. S. Villase$\tilde{\rm n}$or, Class. Quantum Grav. {\bf 25}, 085002 (2008).

\bibitem{CQG25} J. Cortez, G. A. Mena Marug\'an, and J. M. Velhinho, Class. Quantum Grav. {\bf 25}, 105005 (2008).

\bibitem{CMV8} J. Cortez, G. A. Mena Marug\'an, and J.M. Velhinho, Phys. Rev. D {\bf 81}, 044037 (2010).

\bibitem{PRD79} J. Cortez, G. A. Mena Marug\'an, R. Ser\^odio, and J. M. Velhinho, Phys. Rev. D {\bf 79}, 084040 (2009).

\bibitem{CMV83} J. Cortez, G. A. Mena Marug\'an, J. Olmedo, and J. M. Velhinho, Phys. Rev. D {\bf 83}, 025002 (2011).

\bibitem{JCAP10} J. Cortez, G. A. Mena Marug\'an, and J. M. Velhinho, J. Cosmol. Astropart. Phys. {\bf 10}, 030 (2010).

\bibitem{CMOV28} J. Cortez, G. A. Mena Marug\'an, J. Olmedo, and J. M. Velhinho, Class. Quantum Grav. {\bf 28}, 172001 (2011).

\bibitem{gr-qc} J. Cortez, G. A. Mena Marug\'an, J. Olmedo, and J. M. Velhinho, Phys. Rev. D {\bf 86}, 104003 (2012).

\bibitem{brocker} T. Br\"ocker and T. tom Dieck, {\it Representations of Compact Lie Groups} (Springer-Verlag, New York, U.S.A., 1985).

\bibitem{3torus} L. Castell\'o Gomar, J. Cortez, D. Mart\'in-de Blas, G. A. Mena Marug\'an, and J. M. Velhinho, J. Cosmol. Astropart. Phys. {\bf 11}, 001 (2012).

\bibitem{PRL49} A. H. Guth and S. Y. Pi, Phys. Rev. Lett. {\bf 49}, 1110 (1982).

\bibitem{LR314} D. H. Lyth and A. Riotto, Phys. Rept. {\bf 314}, 1 (1999).

\bibitem{hr-sh} R. Honegger and A. Rieckers, J. Math. Phys. {\bf 37}, 4292 (1996).

\bibitem{Luzin} A. N. Kolmogorov and S. V. Fomin, {\it Elements of the Theory of Functions and Functional Analysis} (Dover, New York, U.S.A., 1985).

\bibitem{mukhanov2} V. F. Mukhanov, H. A. Feldman, and R. H. Bradenberger, Phys. Rep. {\bf 215}, 203 (1992).

\bibitem{bardeen} J. M. Bardeen, Phys. Rev. D {\bf 22}, 1882 (1983).

\end{thebibliography}
\end{document}